\documentclass[iop]{emulateapj}
\usepackage{amsmath}

\newcommand{\kms}{\rm ~km~s^{-1}}
\newcommand{\ergs}{\rm ~erg~s^{-1}}

\begin{document}

\title{PARTICLE TRANSPORT IN YOUNG PULSAR WIND NEBULAE}
\author{Xiaping Tang and Roger A. Chevalier}
\affil{Department of Astronomy, University of Virginia, P.O. Box 400325, \\
Charlottesville, VA 22904-4325; xt5ur@virginia.edu, rac5x@virginia.edu}

\begin{abstract}
The model for pulsar wind nebulae (PWNe) as the result of the magnetohydrodynamic (MHD) downstream flow from a shocked, relativistic pulsar wind has been successful in reproducing many features of the nebulae observed close to
the central pulsars.
However, observations of well-studied young nebulae like the Crab Nebula, 3C 58, and G21.5--0.9 do not
show the toroidal magnetic field on a larger scale that might be expected in the MHD flow model;
in addition, the radial variation of spectral index due to synchrotron losses is smoother  than expected in the MHD flow model.
We find that pure diffusion models can reproduce the basic data on nebular size and
spectral index variation for the Crab, 3C 58, and G21.5--0.9.
Most of our models use an energy independent diffusion coefficient; power law variations of
the coefficient with energy are degenerate with variation in the input particle energy
distribution index in the steady state, transmitting boundary case.
Energy dependent diffusion is a possible reason for the smaller diffusion coefficient
inferred for the Crab.
Monte Carlo simulations of the particle transport  allowing for advection and diffusion
of particles
suggest that diffusion dominates over
much of the total nebular volume of the Crab.
Advection dominates close to the  pulsar and is likely to play a role in the X-ray half-light
radius.
The source of diffusion and mixing of particles is uncertain, but may be related to the
Rayleigh-Taylor instability at the outer boundary of a young PWN or to
instabilities in the toroidal magnetic field structure.

\end{abstract}


\section{INTRODUCTION}

The finding of the diminishing size of the Crab synchrotron nebula with increasing frequency
supports the picture that energetic electrons are injected in the vicinity of the pulsar
and lose energy to synchrotron radiation in the larger nebula \citep{wilson72,rees74}.
The radial emission profiles were first modeled as a central particle source with diffusion
into the larger nebula \citep{gratton72,wilson72}.
\cite{rees74} specified a termination shock in the pulsar wind as the source of the
particle acceleration and viewed the outer part of the nebula as a place where the
pulsar magnetic field winds up and the flow decelerates.
This view was put on a firmer basis by \citet[][hereafter KC]{kennel84a,kennel84b}, who calculated the
conditions at the relativistic MHD shock and followed the time independent downstream
flow of fields and particles.
This 1-dimensional model with advected particles was able to reproduce the observed
sizes of the optical and X-ray emission in the Crab Nebula, but did not address the
radio emission.

High resolution imaging of the Crab with the {\it Hubble} Telescope 
at optical wavelengths and {\it Chandra}
at X-ray wavelengths shows an active system of toroidal filaments and jets
close to the pulsar in line with the expected position of the
termination shock \citep{hester08}.
These observations have motivated 2-dimensional MHD simulations, which allow for a 
polar angle  dependence of the pulsar wind power \citep{komissarov03,delzanna04}.
In these models, the wind is stronger in the equatorial plane, producing the 
toroidal filaments.
Hoop stresses in the shocked flow bring material back to the axis to form
the jets.
Current models for the filaments are able to reproduce many aspects of the
filaments \citep[see][for a review]{buccian10},
including the integrated spectrum of the Crab
from radio to TeV \citep{vopi08} and the time variability of the inner
structure \citep{vopi08,camus09,Komissarov11}.
The jet-torus structure near the pulsar has been commonly observed in 
X-ray images of pulsar nebulae \citep{karg08},
and the structure is presumed to be a standard feature of pulsar nebulae.
Although models for the toroidal filaments are now convincing, the nature
of the flow beyond the filaments to the edge of the nebula remains uncertain.

The primary way of exploring the particle transport is to model the structure
of the nebulae at different photon energies or, equivalently, the structure and
photon index distribution.
The crucial point is that the particles lose energy to synchrotron radiation as they age so
that the photon index distribution provides a good test of the particle transport mechanism.
Provided the magnetic field is not strongly varying, the spectral index in
a particular location gives
information on the mean age of the particles.
Observations of the Crab Nebula at optical \citep{veron93} and infrared \citep{Temim06}
wavelengths show a monotonic change in spectral index from the center to the edge
of the nebula, where edge is defined by the decrease in surface brightness at that particular
wavelength.
The well-known bays in the Crab are asymmetric structures, but the spectral index
at the bay edge is similar to that at other edges.
The data do not indicate a highly asymmetric flow in the nebula.
Thus, we assume spherical symmetry in our models.

In Section~2 of this paper, we discuss issues with existing models for the larger scale flow
and, in Section~3, consider a  diffusion/advection model for the particle propagation.
The model is applied to the phase when the pulsar  wind nebula (hereafter PWN) is expanding into
the freely expanding supernova gas, before a reverse shock front moves in
due to interaction with the surroundings.
We concentrate on comparing our models to the Crab Nebula, 3C 58 and
G21.5--0.9 because they are the best observed PWNe within that phase.
In Section~4, we discuss the diffusion process.

\section{OUTER STRUCTURE OF YOUNG PULSAR NEBULAE}

Current data on the Crab Nebula convincingly show that there is a relativistic
flow in the interior that must slow to match the outer boundary \citep{hester08}.
The issue treated in this paper is how the particles are transported between
the inner, toroidal region and the outer boundary of the PWN.
In the KC model, the particles are advected with a toroidal magnetic field.
Cross field scattering of particles is expected to be small \citep[e.g.,][]{dejager08}, so
that diffusion of particles can be neglected.
Problems with this model in reproducing the spectral index distributions in
young PWNe have been raised \citep{reynolds03,slane04}.
This issue is discussed in detail in the next section.
Here we note some other points relevant to the toroidal field model for the outer structure.

 In the Crab Nebula, the X-ray emission is  from a region close to
the pulsar because of synchrotron burn-off.
If one goes to optical wavelengths, where the particles have longer lifetimes,
an analysis of the polarization shows that there are $3-6$ magnetic
elements across the nebula with possible smaller scale structure \citep{felten74}.
\cite{schmidt79} find that magnetic structure in the Crab only extends down to
about 20 arcsec, or 0.2 pc.
\cite{seward06} presented deep {\it Chandra} images of the Crab, finding evidence for fingers
with a roughly radial orientation; the spectral index structure implied rapid diffusion along
the structures, presumably oriented along the magnetic field and slow diffusion perpendicular
to the fingers.
The X-ray emitting particles in 3C 58 have longer lifetimes and imaging X-ray studies with
{\it Chandra} show many magnetic loops without a clear
toroidal structure, except close to 
the pulsar \citep{slane04}; the X-ray filaments are related to ones observed
at radio wavelengths and with some optical filaments (thermal gas).
Overall, there is little evidence for toroidal structure in pulsar nebulae
except near the central pulsars.
Two PWNe that do show evidence for toroidal field structure are the small nebula
around the Vela pulsar \citep{dodson03} and G106.6+2.9 \citep{kothes06}; however, these objects
are probably in a different evolutionary stage than the Crab, without an unstable outer
boundary \citep{chevalier11}.

The Crab Nebula, with radio spectral index $\beta=0.299\pm 0.009$ 
\citep{baars77} where flux $\propto \nu^{-\beta}$, shows little radio spectral index variation over
the entire nebula, to within 0.01 \citep{bieten97}, although PWNe themselves show a range
of spectral indices, e.g., 3C 58 has $\beta=0.07\pm 0.05$ \citep{bieten01}.
There is no indication that the spectral index observed in the Crab is a universal
value, so the uniformity of spectral index is surprising.
G21.5--0.9 also has a fairly uniform radio spectral index image \citep{bieten08},
as well as 3C 58 \citep{bieten01}.

These observational considerations support the view that, although there is clearly toroidal
structure where the pulsar wind impacts the larger nebula, the flow is more
radial in the outer nebula and there is evidence for a mixing process.
There are several possible reasons for the apparent mixing of energetic particles.
The acceleration of the supernova ejecta by the pulsar bubble is
Rayleigh-Taylor unstable \citep{chev77,jun98,hester96}.
The structure observed in the thermal gas filaments in the Crab Nebula
and other PWNe is likely due to this instability \citep{hester08},
although there are still uncertainties about how the instability operates when
the low density fluid is magnetized \citep{buccian04,stone07}.
As discussed by \cite{hester95}, there is evidence in the Crab for magnetic
field lines being `draped' around thermal filaments and stretched in the
radial direction.
The Crab filaments cover the velocity range $700-1500\kms$ \citep{hester08}.

Another possibility is the action of instabilities occurring at or near the
pulsar wind termination shock \citep{begel98,camus09,mizuno11}.
Instabilities right at the shock front may not be compatible with the apparent
regular structure observed in the case of the newly formed nebula
around the Vela pulsar \citep{dodson03,chevalier11}.
There may be feedback between instabilities near the termination shock and the
outer boundary.
 In their axisymmetric numerical simulations, \cite{camus09} find that waves and vortices
in the larger nebula feed back on the structure at the termination shock,
which in turn generates more structure in the nebula.
\cite{camus09}  note that it
is impossible to distinguish between the cause and the effect.
In addition, some thermal gas from the supernova is entrained in the unstable
region, explaining the optical filaments seen in association with nonthermal
filaments in the Crab and 3C 58.

These observational and theoretical considerations show that, although the region close to the pulsar has a clear toroidal structure, the larger nebula has a complex structure that includes 
a radial component to the magnetic field.
As summarized by \cite{hester08}, there are layers of magnetic fields folded on top
of each other such that adjoining field lines in one place move away from each other.
Although diffusion across magnetic field lines is expected to be small, even a small amount
of cross field transport could result in thorough mixing, with this magnetic field configuration.
We thus consider models with radial diffusion.

One uncertainty for the models is the degree to which particles are transmitted through
the outer boundary of the PWN.
The expectation is that the PWN magnetic field is contained within the wind bubble, which is
bounded by supernova ejecta in young PWNe; the outer boundary of the Crab Nebula shows loop
structures at radio wavelengths that are likely to delineate the magnetic field \citep{hester08}.
The characteristic mean free path for particles to cross the magnetic field is limited to
the particle gyroradius (Bohm limit)
\begin{equation}
\lambda=10^{-4}\left(\frac{E_e}{100{\rm~TeV}}\right)\left(\frac{B}{100{\rm~\mu G}}\right)^{-1}{\rm~pc,}
\end{equation}
where the  particle energy, $E_e$, corresponds to an energy of synchrotron radiation
through $E_e= (20 {\rm~TeV})(B/100{\rm~\mu G})^{-1/2}E_{keV}^{1/2}$
and $B$ is the magnetic field.
The escape time for Bohm diffusion from a region of size $R$ is then \citep{dejager08}
\begin{eqnarray}
t_{esc}\approx 16,000 \left(\frac{R}{2{\rm~pc}}\right)^{2}\left(\frac{E_e}{100{\rm~TeV}}\right)^{-1} \nonumber\\
\times\left(\frac{B}{100{\rm~\mu G}}\right){\rm~yr,}
\end{eqnarray}
which is long compared to the ages of the PWNe considered here ($\sim 10^3$ yr),
particularly for particles radiating below X-ray wavelengths.
However, there is some chance for escape from a narrow region close to the edge
of the nebula.

An observational test for the transmission of particles would be synchrotron emission
from energetic particles that have left the main PWN.
There has been little evidence for such emission, but \cite{bamba10} have recently 
found X-ray emission around a number of PWNe,
which they interpret as synchrotron
emission from escaped particles.
To have X-ray emitting particles extend to such large radial distances requires a
surprisingly low magnetic field strength in order to avoid synchrotron losses.
One of the PWNe discussed by \cite{bamba10} is G21.5--0.9, which is also one of
the primary remnants treated here.
However, there are other interpretations for the extended emission.
\cite{bocchino05} attributed the emission to a combination of a dust scattering halo
plus emission from a surrounding shell;
\cite{matheson10} have shown clear evidence for shell emission.

Escape of electrons and positrons from PWNe has also come up in the context of
a possible source for features in the Galactic cosmic ray spectrum of electrons
and positrons \citep[e.g.,][]{chang08}.
However, the escape can be from elderly PWNe ($>10^5$ yr old), and does not
require escape from young objects like those discussed here \citep[e.g.,][]{malyshev09}.
\cite{hinton11} suggested the escape of particles from the Vela X PWN to explain
the steeper particle spectrum in the outer parts of the nebula; however, this is an
older nebula that has likely been affected by the reverse shock wave \citep{blondin01}.
Overall, particle escape in PWNe with ages $\sim 10^3$ yr does not appear likely.
In Section 3.2, we model the effect of the outer boundary condition for the PWN.

In the early diffusion models for the Crab Nebula \citep{gratton72,wilson72,weinberg76},
the diffusion coefficient was assumed to be constant with energy, and that is the assumption
that we make in most of our modeling.
\cite{weinberg76} argued for a constant coefficient based on the fact that the diffusion length
is likely to be related to the size of magnetic filaments and  is much larger than the
gyroradius.
However, there are reasons to consider energy dependent diffusion of the form $D_E(E)=D_0E^\alpha$,
where $\alpha$ is a constant.
On the observational side, cosmic rays are known to diffuse from the Galaxy in an energy dependent
way \citep[e.g.,][and references therein]{strong07}.
The data indicate $\alpha=0.3-0.6$, with a diffusion coefficient of $(3-5)\times 10^{28}$ cm$^2$
s$^{-1}$ at a reference particle energy of 1 GeV \citep{strong07}.
Interstellar turbulence is thought to play a role in the energy dependent diffusion.
Also, there is the possibility that the diffusion coefficient is proportional to
the particle gyroradius, as in Bohm diffusion, leading to $\alpha=1$.
This ``Bohm-type'' value of $\alpha$ is used by \cite{etten11} 
and \cite{hinton11} in their modeling of  evolved
PWNe.
However, the diffusion length is much larger than the particle gyroradius, so there is
not a clear argument for $\alpha=1$ in the PWN case.
We consider the possible effect of energy dependent diffusion in Section 3.1.

\section{MODELS WITH DIFFUSION}

\subsection{Pure diffusion model}

\cite{wilson72} first showed that the spatial and spectral distribution of the Crab Nebula in the optical could be explained by a  diffusion model with synchrotron radiation losses. 
Observations of 3C 58 \citep{bocchino01,slane04} and G21.5-0.9 \citep{slane00,safi01} taken by {\it Chandra} also gave a photon index distribution that is similar to the optical spectral index
distribution in the Crab and is incompatible with the KC model \citep{reynolds03,slane04}. 
Although observations of the Crab clearly show evidence for a relativistic wind close
to the pulsar, a pure diffusion model illustrates one limiting case of the expected
particle transport.
We used the pure diffusion model developed by  \cite{gratton72}  to fit the spectral index distribution of Crab from radio to optical, and 3C 58 and G21.5-0.9 in X-rays. We also used the model to calculate the half light radius of the Crab from radio to X-rays. 
This is the same model used by \cite{wilson72} in his model for the Crab.
The model assumes that a point source injects particles into an infinite space with spherical symmetry, and the injected particles follow a power law distribution $N(E,r=0)=KE^{-p}$. The transport
mechanism for the injected particles is diffusion. In order to satisfy the spherical symmetry assumption,  the objects we discuss in this paper are young PWNe that are observed
to have  approximate spherical symmetry. In our initial model we have a pre-defined PWN radius $R$; when we calculate the half light radius and integrated spectrum, we only consider the particles that are within the nebular radius $R$, reasoning that the magnetic field in the freely expanding ejecta
outside $R$ is small.  The outer nebular boundary is a transmitting boundary.
The model also 
assumes that the diffusion coefficient $D$ and magnetic field $B$ inside $R$ are constant with radius. Here we neglect the adiabatic expansion energy losses and assume synchrotron radiation loss is the only energy loss, so that $dE/dt=-QE^2$, where $Q=2.37\times10^{-3}B_\perp^2\ergs$ in cgs units and $B^2_\perp=(2/3)B^2$ \citep{pachol70}.    We examine the assumptions of point source injection, pure diffusion, and a transmitting boundary in Section 3.2. In this section we use this simple pure diffusion model to analyze the spectral index distribution and half light radius of PWNe.

Based on these assumptions, the  number density distribution $N(E,r,t)$ is \citep{gratton72}
\begin{equation}
N(E,r,t)=\frac{K}{4\pi r D}E^{-p}f_p (u,v),
\label{numdistr}
\end{equation}
where
$$
u=\frac{r^2QE}{4D}
$$
and
$$
f_p(u,v)=\frac{1}{\sqrt{\pi}}\int^\infty_v
\left(1-\frac{u}{x}\right)^{p-2}\frac{e^{-x}}{\sqrt{x}}dx.
$$
The lower limit $v$ of the integral $f_p(u,v)$ is
 \[v = \left\{ 
\begin{array}{l l}
  \frac{r^2}{4Dt} & \quad \mbox{if $t < \frac{1}{QE}$}\\
  \frac{r^2QE}{4D} & \quad \mbox{if $t > \frac{1}{QE}$}\\ \end{array} \right. \] 
if there is no upper limit for the injection particle energy. 
Here $t$ is the age of the nebula.

 In order to simplify the calculation of the emission, we further assume that all the radiated power $P$ of an electron of energy $E$ goes into radiation of a frequency $\nu$ corresponding to the maximum synchrotron radiation power. Therefore 
\begin{equation} 
P(\nu)=C (B_{\perp} \nu)^{1/2} N[E(\nu)], 
\end{equation}
where $C$ is a constant, and  
\begin{equation}
E(\nu)=\left(\frac{4\pi m_e^3c^5}{0.29\times 3e}\right)^{1/2}\left(\frac{\nu}{B\sin\theta}\right)^{1/2},
\end{equation}
where $m_e$ and $e$ are the mass and charge of an electron, and $\theta$ is the particle
pitch angle.
Here we assume $B\sin\theta=B_{\perp}=(2/3)^{1/2}B$, yielding $E(\nu)=7.42\times 10^{-10}(\nu/B_{\perp})^{1/2}$ erg.
The spectral index distribution $S(r)$ between frequency $\nu_1$ and $\nu_2$ is given by
\begin{eqnarray}
\lefteqn{S(r)=\frac{\log[P_{\nu1}(r)/P_{\nu2}(r)]}{\log(\nu_1/\nu_2)}} \nonumber \\ 
 & & =\frac{\log(\nu_1/\nu_2)^{1/2}+\log[N_{tot}(\nu_1,r)/N_{tot}(\nu_2,r)]}{\log(\nu_1/\nu_2)},
\end{eqnarray}
where $N_{tot}(\nu,r)$ is the total number of particles emitted per unit area per unit time, per unit frequency  with frequency $\nu$ and at radius $r$ from a central point source after integration along the line of sight.  \cite{gratton72} assumed a point source which makes  
$r=0$ a singularity in $N_{tot}(\nu,r)$.
We performed integrations along the line of sight  starting at a cutoff radius, but this did not affect the larger scale results.

There is a critical energy for synchrotron cooling, $E_{crit}=1/Qt$, 
which is relevant for the number density distribution $N(E,r,t)$. If $E>E_{crit}$, $N(E,r,t)$ reaches a steady state solution $N(E,r)$. 
If $E<E_{crit}$, only particles with $E_{initial}<E/(1-QEt)$ contribute to the spectrum at the frequency
$\nu$ and $N(E,r,t)$ evolves with time. The corresponding peak synchrotron radiation frequency of particle with $E_{crit}$ is  
\begin{equation}
\nu_{crit}=6\times 10^{14} \left(\frac{10^3{\rm~yr}}{t}\right)^2 
\left(\frac{100~\mu{\rm G}}{B}\right)^3 {\rm~Hz}.
\label{critv}
\end{equation}
If the injected particles have an upper limit of energy $E^*$, the lower limit $v$ of the integral $f_p(u,v)$  changes to 
 \[v = \left\{ 
\begin{array}{l l} 
  \frac{r^2}{4Dt} & \quad \mbox{if $t < \frac{1}{Q}\left(\frac{1}{E}-\frac{1}{E^*}\right)$}\\
  \frac{r^2Q}{4D\left(\frac{1}{E}-\frac{1}{E^*}\right)} & \quad 
\mbox{if $t > \frac{1}{Q}\left(\frac{1}{E}-\frac{1}{E^*}\right)$}\\ \end{array} \right. \] 
and the critical energy becomes $E^*/(1+QtE^*)$. 
When the energy range of interest satisfies  $E \ll E^*$, the $E^*$ term in $v$ is not important and can be neglected, which means we can assume the injection particle energy has no upper limit. A plausible estimate for $E^*$ is that the gyroradius of the particle equals to the termination shock radius, $R_s$, or
$R_s=R_{gyro}=E/eB$
for a relativistic electron. 
 We then have $E^*=R_seB$ and the corresponding peak synchrotron emission frequency is 
$$\nu^*=3.3\times 10^{22}\left(\frac{R_s}{0.1{\rm pc}}\right)^2\left(\frac{B}{100\mu{\rm G}}\right)^3 {\rm~Hz}.$$
For the objects considered in this paper, electron energy injection with no upper limit is always a good assumption for frequencies below soft X-rays. 

The spectral index distribution of the system at a certain frequency band mainly depends on the ratio $\eta$ between diffusion distance $d=(6Dt)^{1/2}$ and the nebular size $R$. At a certain frequency, the same $\eta=d/R=(6Dt/R^2)^{1/2}$  gives roughly the same spectral index distribution. The $p$ index of the injected particles would also affect the spectral index profile to some extent. Since we  have  good observational data for the spectral index of the PWNe considered here, its value can be directly determined from the observations. If the frequency band is in the steady state regime, then $t=1/QE$, and we find $\eta\propto (D/\nu^{1/2} B^{3/2}R^2)^{1/2} $.
Since the nebular size is usually known, the spectral index distribution is determined by $\eta\propto (D/B^{3/2})^{1/2}$ in a particular frequency band. The diffusion coefficient $D$ and magnetic field $B$ are coupled together in the spectral index fitting; in order to get an accurate 
diffusion coefficient $D$, we need to know the magnetic field $B$ of the system. One way to estimate  $B$ is based on the synchrotron break frequency  in the integrated spectrum
(equation [\ref{critv}]). 
However, the break is not a sharp feature and it is difficult to locate the synchrotron break frequency of the Crab and 3C 58 based on current observational data \citep{slane08,arendt11}. Another way is to model the high energy inverse Compton emission. 
The magnetic field obtained by using inverse Compton fitting is slightly different from the magnetic field defined in our pure diffusion model, because in our model we solve for the constant $B$ situation which means the magnetic field $B$  is an average $B$ of the PWN over its lifetime. Our average magnetic field should be slightly larger than the magnetic field indicated by inverse Compton modeling. The minimum energy method for synchrotron emission  also gives an estimate for  $B$,                                      
but it depends on an uncertain assumption. 
If the frequency band is in the non-steady state regime, $\eta=d/R=(6Dt/R^2)^{1/2}$, where $t$ is now  the age of the nebula. The diffusion coefficient $D$ and magnetic field $B$ are now decoupled and $\eta\propto (D)^{1/2}$. 
However, in this regime the particles do not suffer synchrotron losses, so their spectrum is not changed from the injection spectrum and they do not give  useful information about the diffusion coefficient.

\begin{figure}
        \epsscale{1.0} \plotone{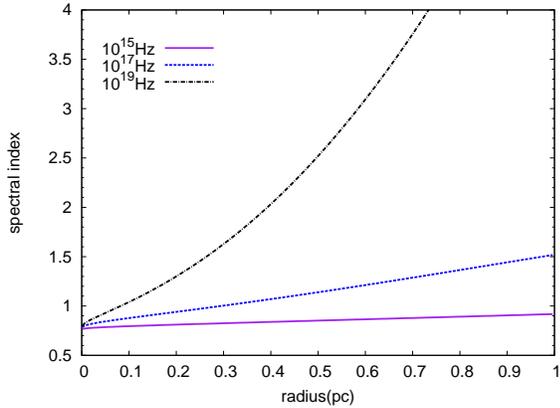}
        \caption{\label{3C_58_vary_freq} Flux spectral index ($\beta$) distribution of a PWN with $p=2.5$, nebular radius $R=1$ pc, $B=100$ $\mu$G, and $D=10^{27}$ cm$^2$ s$^{-1}$ in a pure diffusion model.
}
\end{figure}

Next we consider how the spectral index profile varies as a function of the ratio $\eta$. As we are mainly interested in the steady state case, we now consider the critical frequency $\nu_R$ corresponding to the case $\eta\approx (4Dt)^{1/2}/R=(4D/QER^2)^{1/2}=1$. It is the same as $\nu_B$ defined in \cite{gratton72}:
\begin{eqnarray}
\nu_R=1\times 10^{17} \left(\frac{D}{10^{27}{\rm~cm^2~s^{-1}}}\right)^2\nonumber\\
\times\left(\frac{1{\rm ~pc}}{R}\right)^4 \left(\frac{100~\mu
{\rm G}}{B}\right)^3  {\rm~Hz}.
\label {critR}
\end{eqnarray}
We  assume a PWN in steady state with $p=2.5$, and scale to a nebular size $R=1$ pc, magnetic field $B=100$ $\mu$G and diffusion coefficient $D=10^{27}$ cm$^2$ s$^{-1}$.  We calculate the spectral index distribution for the three cases: $\nu \ll \nu_{R}, \nu \approx \nu_{R}$, and $\nu \gg \nu_{R}$ (Figure \ref{3C_58_vary_freq}).  When $\nu \ll \nu_{R}$, which corresponds to $\eta \gg 1$, the photon index profile is flat. At high frequency the photon index profile first changes into a power law when $\nu \approx \nu_{R}, \eta\approx 1$, and then  into an exponential  when $\nu \gg \nu_{R}, \eta \ll 1$. When the diffusion distance $d \gg$ $R$, the diffusing particles within the nebula are well mixed so the spectral index profile tends to be flat. When the diffusion distance $d<R$, the  particle density  drops quickly along the radial direction because of the short cooling time, so the spectral index  shows steepening in the radial direction. 

\begin{figure}
        \epsscale{1.0} \plotone{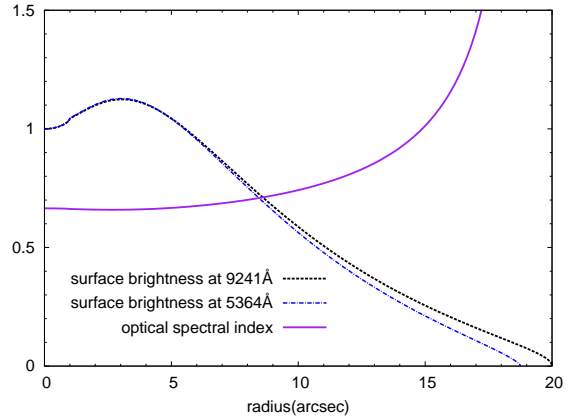}
        \caption{\label{KC_model} Flux spectral index ($\beta$) distribution of the Crab at optical wavelengths, based on the model of \cite{kennel84b}.   The surface brightness is normalized to the value at the center of Crab.
}
\end{figure}

A calculation of the spectral index distribution in the KC model  shows that it has a problem in explaining the observed spectral index profile \citep[see also][]{reynolds03}. We take the Crab as an example and use the KC model to calculate the spectral index distribution of the Crab at optical wavelengths, where the KC model is still applicable. We use the best fit parameters given by \cite{kennel84b} to do the calculation and assume that there is no synchrotron emission within the termination shock. In order to give a good comparison, we use the same emissivity as used in \cite{kennel84b}, which is slighty different from our value. We add a pre-defined radius $R$ which is 20 times of the termination shock radius in the simulation; there is no emission beyond this point. The results are shown in Figure \ref{KC_model}. The spectral index profile in Figure \ref{KC_model} does not  fit the optical data shown in \cite{veron93}. In the observations, the spectral index profile is approximately a power law distribution, while the results given by the KC model are  flat within a certain radius and then increase very quickly beyond that radius. The power law like spectral index distributions are also seen 
at X-ray wavelengths in 3C 58 \citep{slane04} and G21.5--0.9 \citep{slane00,safi01} which indicates that diffusion processes could be generally important in young PWNe \citep{reynolds03}. \cite{delzanna06} show that a 2D MHD simulation could reproduce most of the toroidal and jet like structure near the termination shock, but the spectral index properties of the Crab Nebula suggest diffusion processes on larger scales.    

The nebular size of a PWN is also determined by $\nu_R$ or $\eta$. 
 When $\nu < \nu_R$, $\eta>1$ in the steady state case, the nebular size remains the same due to the boundary condition. When $\nu > \nu_R$, $\eta<1$, the  size tends to shrink as the cooling time of particles is smaller than the diffusion time. In the $\nu > \nu_R$, $\eta<1$ regime, the nebular size for the pure 
diffusion model can be estimated by setting $\eta=1$, which yields $R\approx(6D/QE)^{1/2}$. 

\begin{figure}
        \epsscale{1.0} \plotone{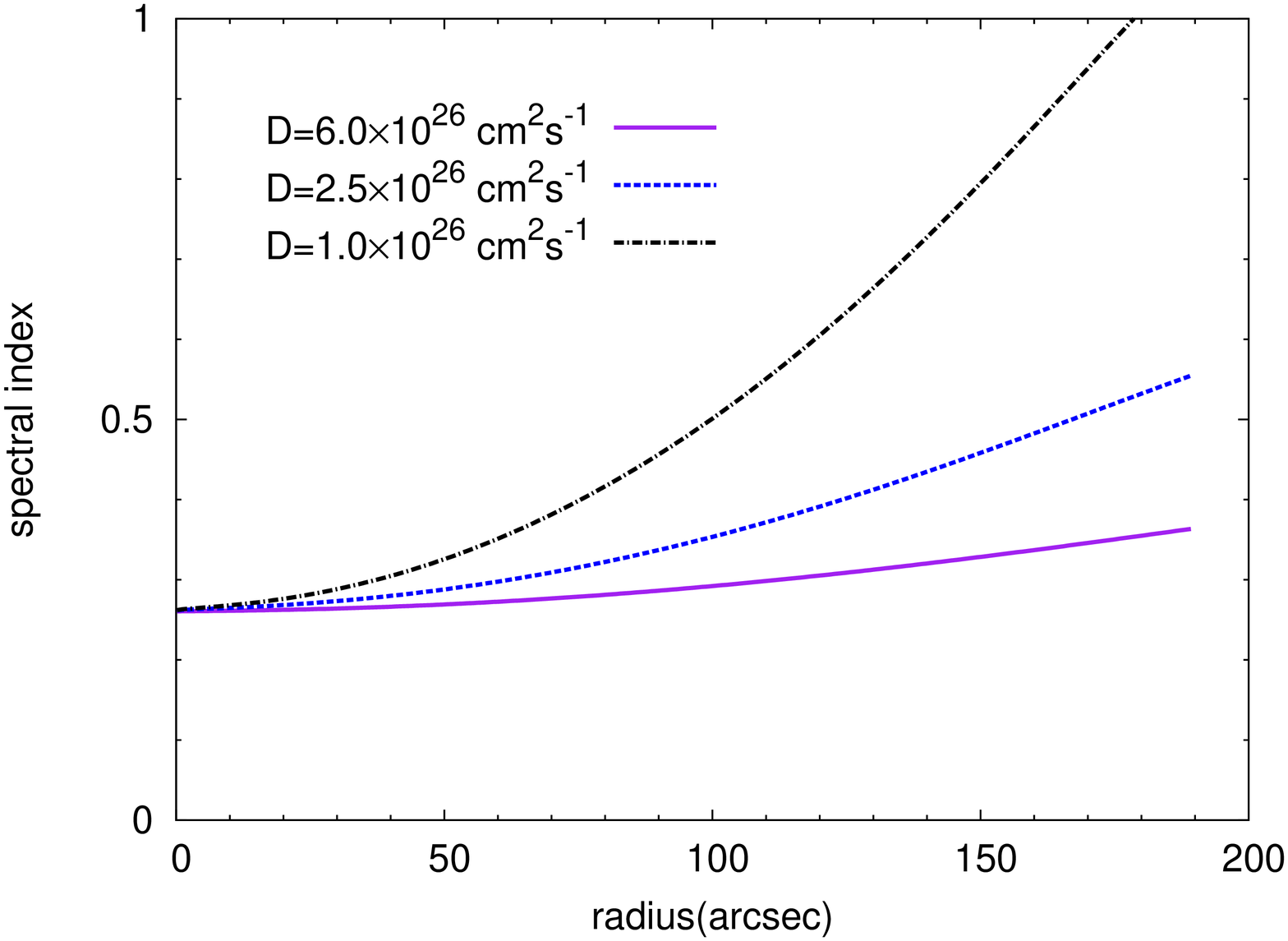}
        \caption{\label{crab} Flux spectral index ($\beta$) distribution of the Crab from 5 GHz to $6\times 10^{14}$ Hz, assuming $p=1.52$ and $B=300$ $\mu$G in a pure diffusion model.
}
\end{figure}

\begin{figure}
        \epsscale{1.0} \plotone{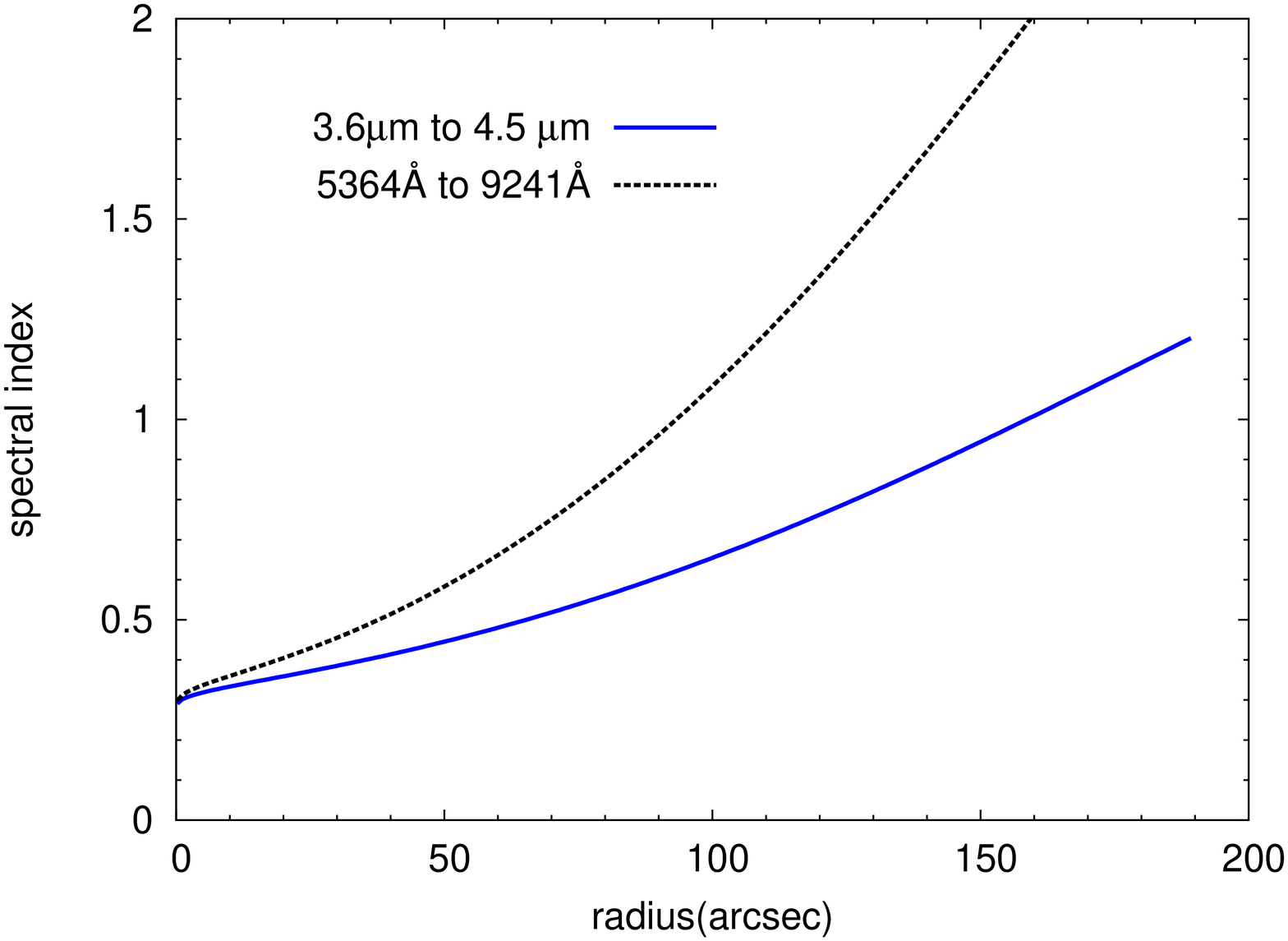}
        \caption{\label{crabfig} Flux spectral index ($\beta$) distribution of the Crab at IR and optical wavelengths, assuming $p=1.52$, $B=300$ $\mu$G, and $D=2.5\times 10^{26}$ cm$^2$ s$^{-1}$ in a pure diffusion model,
}
\end{figure}

We first use parameters known from observations such as age $t$, nebular size $R$ and magnetic field $B$ to fit the spectral index distribution of the Crab, 3C 58 and G21.5--0.9.  Fitting a model
yields the diffusion coefficient $D$ and $p$ of the PWNe. 
We then discuss the nebular size behavior.
For the Crab we use a magnetic field $B=300$ $\mu$G.  This is slightly larger than the value, $\sim 200$ $\mu$G, found by \cite{deJager1996} and \cite{Aharonian2004} from inverse Compton emission, which gives the current value of the field. Our value gives a sufficiently high  diffusion coefficient $D$ to explain the size of the Crab. 
Radio data for the Crab show $p=1.52$ .
The spectral index distribution from radio (5 GHz) to optical  $6\times 10^{14}$ Hz frequencies is shown in Figure \ref{crab}. By comparing our results with the major axis data in \cite{wilson72} we find that $D=2.5\times 10^{26}$ cm$^2$ s$^{-1}$ gives a good fit to the data. Since we do not know exactly the optical frequency  used in \cite{wilson72}, we did not
do a least squares fit.  We then used the diffusion coefficient $D=2.5\times 10^{26}$ cm$^2$ s$^{-1}$ to calculate the spectral index distribution for infrared ($3.6-4.5$ $\mu$m) and optical ($5364-9241$ \AA) wavelengths, which are shown in Figures \ref{crabfig}. In the infrared (IR), fitting the spectral index  variation from 0.3 to 0.8 within the Crab nebula found by \cite{Temim06} requires the nebular radius to be  $\sim130^{\prime\prime}$.  Our simulation  indicates that in the IR the nebular size of the Crab has  decreased due to synchrotron losses, which is consistent with  Figure 3 of \cite{Temim06}.
The spectral index variation from 0.6 to 1.1 within the Crab Nebula at optical wavelengths \citep{veron93} implies the nebular size of the Crab  is  $\sim100^{\prime\prime}$, which is consistent with the results in Figure 2 of \cite{amato00}.  Comparing our simulation results in Figures \ref{crab} and  \ref{crabfig} with the data in \cite{wilson72} and \cite{veron93} shows that our spectral index in the central region is lower than the observed value if the optical band is involved. We attribute the discrepancy to an intrinsic break in the injected particle spectrum, which is discussed below. The main uncertainty in the model comes from the magnetic field $B$, although other factors, such as non-spherical symmetry and boundary conditions,  also have some effect.     

\begin{figure}
        \epsscale{1.0} \plotone{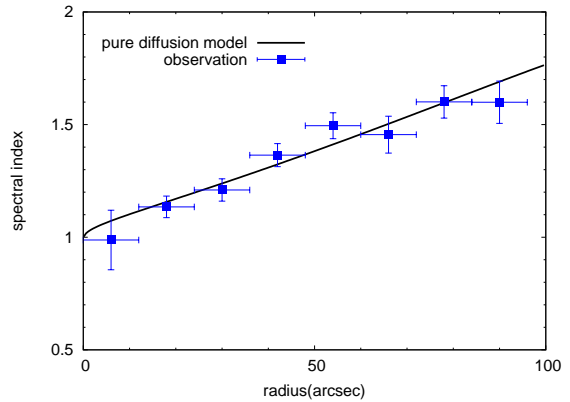}
        \caption{\label{3C_58} Flux spectral index ($\beta$) distribution of 3C 58 from 2.2 keV to 8 keV, assuming $p=2.93$, $B=80$ $\mu$G and
$D=6.11\times 10^{27}$ cm$^2$ s$^{-1}$ in a pure diffusion model.
}
\end{figure}

\begin{figure}
        \epsscale{1.0} \plotone{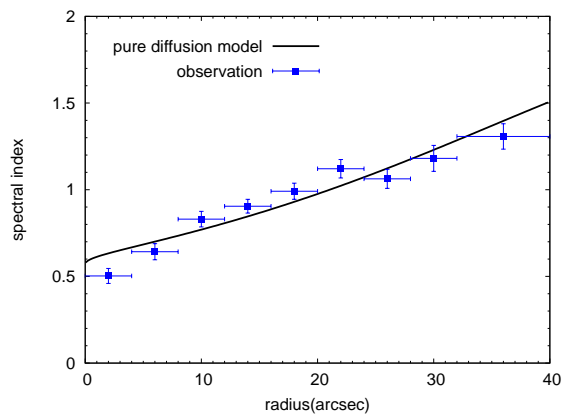}
        \caption{\label{G21.5_0.9} Flux spectral index ($\beta$) distribution of G21.5-0.9 from 0.5 keV to 10 keV assuming $p=2.08$, $B=180$ $\mu$G and $D=3.66\times 10^{27}$ cm$^2$ s$^{-1}$ in a pure diffusion model.
}
\end{figure}

We used the same formalism to fit the X-ray photon index profiles of 3C 58 and G21.5--0.9.
According to equation (\ref{critv}), the X-ray emitting particles of both 3C 58 and G21.5--0.9 have reached a steady state if they are young PWNe, so their age information is not required for photon index fitting and we can use the steady state solution to calculate the photon index distribution.
The parameters we used for 3C 58 and G21.5--0.9 are  listed in Table \ref{params}.
\begin{table*}[ht]
\centering
\caption{Parameters used for modeling photon index profile}
\begin{tabular}{lccccc}
\hline\hline
 Object &Frequency band&Magnetic field $B$&Distance&Angular size &Age \\
&&($\mu{\rm G}$)&(kpc)& of PWN (arcsec)& (yr)\\ \hline
Crab & $5\times 10^9 - 6 \times 10^{14}$ Hz & 300 & 2.0 & 190& 957\\ \hline
Crab & $3.6 - 4.5$  $\mu$m & 300 & 2.0 & 190& 957\\ \hline
Crab & $5364-9241$ \AA  & 300 & 2.0 & 190& 957\\ \hline
3C 58 & $2.2 -  8$ keV & 80 & 3.2 & 100 &  \\ \hline
G21.5-0.9& $0.5 - 10$ keV & 180 & 5.0 & 40 &  \\ \hline
\end{tabular} 
\label{params}
\end{table*}
For 3C 58, the magnetic field $B=80$ $\mu$G is based on the minimum energy condition \citep{green92}.
For G21.5--0.9, the magnetic field $B=180$ $\mu$G is based on the equipartition condition \citep{safi01}.
We used a least squares method to fit the photon index data of both 3C 58 \citep{slane04} and G21.5--0.9 \citep{slane00}. The 2 parameters in each fit are the injected particle spectral index $p$ and diffusion coefficient $D$.
The best fit for 3C 58 between 2.2 keV and 8 keV with $\chi^2_{red}=0.83$ gives $p=2.93$ and $D=6.1\times 10^{27}$ cm$^2$ s$^{-1}$ (Figure \ref{3C_58}). 
For G21.5--0.9, the best fit between 0.5 keV and 10 keV with $\chi^2_{red}=3.28$ gives $p=2.08$ and $D=3.7\times 10^{27}$ cm$^2$ s$^{-1}$ (Figure \ref{G21.5_0.9}).
The diffusion coefficients $D$ we obtained for 3C 58 and G21.5--0.9 are  higher than for the Crab. Part of the reason may be uncertainties in  the magnetic field $B$ of 3C 58 and G21.5--0.9. As discussed above, the diffusion coefficient $D$ and magnetic field $B$ follow  $D\propto B^{3/2}$. If the magnetic field $B$  is lower than our estimate, the diffusion coefficient $D$  also drops; this would imply
a high particle energy in the nebulae because we used the minimum energy value of $B$.
We found that the value of $p$ is insensitive to $B$ and $D$.

\begin{figure}
        \epsscale{1.0} \plotone{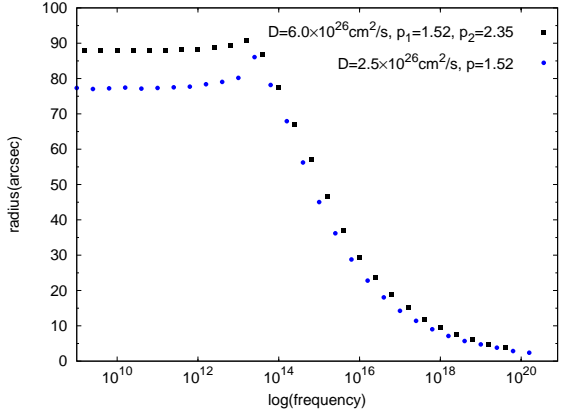}
        \caption{\label{crab_radius} Crab Nebula half-light radius based on a pure diffusion model with $B=300$ $\mu$G.
}
\end{figure}

By using  equation (\ref{critR}), we obtain $\nu_{R}=2\times 10^{13}$ Hz for the Crab if
$D=2.5\times10^{26}$ cm$^2$ s$^{-1}$ and $B=300$ $\mu$G; $\nu_R=1.3\times 10^{18}$ Hz for 3C 58 if
$D=6.1\times 10^{27}$ cm$^2$ s$^{-1}$ and $B=80$ $\mu$G; $\nu_R=2.6\times 10^{17}$ Hz for G21.5--0.9 if $D=3.7\times 10^{27}$ cm$^2$ s$^{-1}$ and $B=180$ $\mu$G. For the Crab, X-ray, optical and near-IR frequencies are all in the
$\nu>\nu_R$ regime, so the nebular size decreases from radio to X-ray.
For 3C 58 and G21.5--0.9, all frequencies below soft X-rays are in the $\nu<\nu_R$ regime, so the
radio, optical and soft X-ray nebular sizes of 3C 58 and G21.5--0.9 tend to be similar. The different 
behavior of nebular size as a function of frequency among the Crab, 3C 58 and G21.5--0.9 is due to the fact that the Crab has a larger magnetic field but lower diffusion coefficient. We use our pure diffusion model to calculate the half light radius of the Crab Nebula with $p=1.52$, $B=300$ $\mu$G and $D=2.5\times 10^{26}$ cm$^2$ s$^{-1}$ (Figure \ref{crab_radius}), assuming that there is an upper limit to the injected particle energy $E$ which corresponds to a frequency of $5\times 10^{22}$ Hz. There is a bump in the half light radius plot,  mainly because for the Crab $p=1.52$, which is $<2$
\citep{pachol70}, and we are assuming synchrotron radiation only emits at the peak frequency. If we consider the full synchrotron spectrum, the bump is diminished. Comparing our results to Figure 2 in \cite{amato00} shows that our model  prediction gives half light radii near the lower limit of radio and optical data for the Crab.  At X-ray wavelengths, our theoretical half light radius is much smaller than the observed value. There are several reasons for this. First, the spherical symmetry assumption breaks down at X-ray wavelengths for the Crab. The {\it Chandra} image of the Crab shows clear toroidal structure near the termination shock \citep{hester08}. Second, in our model we assume a point source while it is in fact an extended source. The termination shock  has an angular size  $\sim 10^{\prime\prime}$ (0.1 pc) at the Crab Nebula. At radio and optical wavelengths, the nebular size is much larger than the size of the injection region so the point source assumption is adequate, but at X-ray wavelengths  it is no longer true.  The last reason is that  our pure diffusion model does not include the effect of advection.  It is likely that both advection and diffusion play a role in  PWNe.  Assuming $V_{adv} \propto r^{-2}$ near the pulsar wind termination shock (KC), $t_{adv}/t_{diff} \propto R $, so advection becomes more important in the inner regions. We expect advection to play some role in X-ray emission from the Crab, and we
discuss it in  Section 3.2. 

In considering the integrated number density $N(E,t)$, we note the result for synchrotron 
losses only \citep{pachol70}
\begin{eqnarray}
\lefteqn{N(E,t)=} \nonumber\\
& & \frac{K}{(p-1)Q}E^{-(p+1)}[1-(1-QEt)^{p-1}], \quad\mbox{if } QEt\leq 1\nonumber\\
& & \frac{K}{(p-1)Q}E^{-(p+1)},\quad\mbox{if } QEt > 1.\nonumber\\
\label{totnumref}
\end{eqnarray}
Our pure diffusion model deviates from this result because we have a pre-defined PWN radius $R$ and assume a transparent outer boundary at $R$; particles that diffuse out of $R$ are not taken into account in the integrated number density $N(E,t)$. The results shown in equation (\ref{totnumref}) would apply to a pure diffusion model with a constant magnetic field $B$ and a reflecting outer boundary which counts all the injection particles. 
As discussed in Section 2, the issue of transmission through the outer boundary of a PWN is
uncertain from the observational point of view.
We discuss the spectral index distribution and half light radius of  a model with a reflecting outer boundary in Section 3.2.

\begin{figure}
        \epsscale{1.0} \plotone{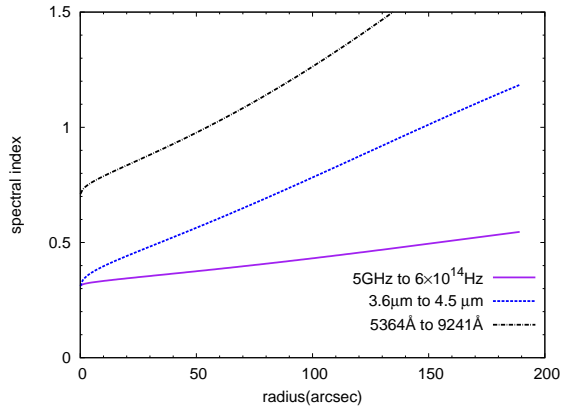}
        \caption{\label{crab_double} Flux spectral index ($\beta$) distribution of the Crab for different wavelength bands assuming $p_1=1.52$, $p_2=2.35$, $B=300$ $\mu$G, and $D=6.0\times 10^{26}$ cm$^2$ s$^{-1}$ in a pure diffusion model.
}
\end{figure}

The Crab, 3C 58 and G21.5--0.9  show flat radio spectra and cannot be fitted by a single power law injection spectrum at radio through X-ray wavelengths even taking into account the evolution of the PWN \citep[e.g.,][]{reynolds84}. 
\cite{buccian11} considered a 1 zone model with a broken power law injection spectrum and long term evolution of the PWN, showing that it can explain the integrated spectra of the Crab and 3C 58 from radio to X-rays. The intrinsic spectral breaks for all PWNe considered are at a similar energy. \cite{sironi11} did both 2D and 3D particle in cell simulation for the termination shock and show that it could create both a flat power law ($p\sim 1.5$) and steep power law ($p\sim 2.5$) components in the post-shock spectrum. The spectral break of PWNe between radio and optical wavelengths may be a natural consequence of particle acceleration at the termination shock. In  fitting the
optical spectral index distribution for the Crab, we  already mentioned that there are indications of another power law component in the Crab. Here we  use a double power law injection spectrum in our pure diffusion model and
re-calculate the spectral index distribution and half light radius of the Crab. The evolution of the PWN is still ignored here because of the additional complications. For 3C 58 and G21.5--0.9, the X-ray spectral index data are well above the spectral break frequency and our modeling is not affected by the double power law feature.  We continue using a magnetic field $B=300$ $\mu$G and $p_1=1.5$ for the low energy component of Crab particle distribution. We use   $p_2=2.35$ and a break energy $4\times10^{11}$ eV for the other power law component as found by \cite{buccian11} and find that $D=6.0\times 10^{26}$ cm$^2$ s$^{-1}$  gives a good fit to the major axis data in \cite{wilson72}, as shown in Figure \ref{crab_double}. Then we use the same diffusion coefficient $D$ to calculate the spectral index distribution at IR and optical wavelengths (Figure \ref{crab_double}). After adding another power law component, we  obtain a better fit to the central spectral index data in \cite{wilson72} and \cite{veron93}. The nebular radius we need to explain the spectral index variation at IR \citep{Temim06} and optical  \citep{veron93} wavelengths becomes smaller: $\sim 110^{\prime\prime}$ in IR and $\sim70^{\prime\prime}$ in optical wavelengths. Our simulation gives a spectral index of 0.7 at the center of the Crab at optical wavelengths, which is slightly larger than the observed value. This is due to the fact that we use $p_2=2.35$ \citep{buccian11} for the steep power law. However, the best fit parameter \cite{buccian11} obtained in their 1D evolution model may not be the best fit parameter for
our case as we are considering a steady-state situation. If we change the break energy for the two power laws, we could obtain a better fit to the optical nebular radius, but the improvement is not significant in view of the other uncertainties in the model.  The half light radius of the Crab with double power law injection spectrum is shown in Figure \ref{crab_radius}.   The double power law fit gives a larger nebular size in the radio band which is in the non-steady  state regime, mainly because we use a larger diffusion coefficient $D$ in the double power law fit. 

\begin{figure*}
\begin{center}
$\begin{array}{cc}
(a)&(b)\\
 \includegraphics[width=3.1in]{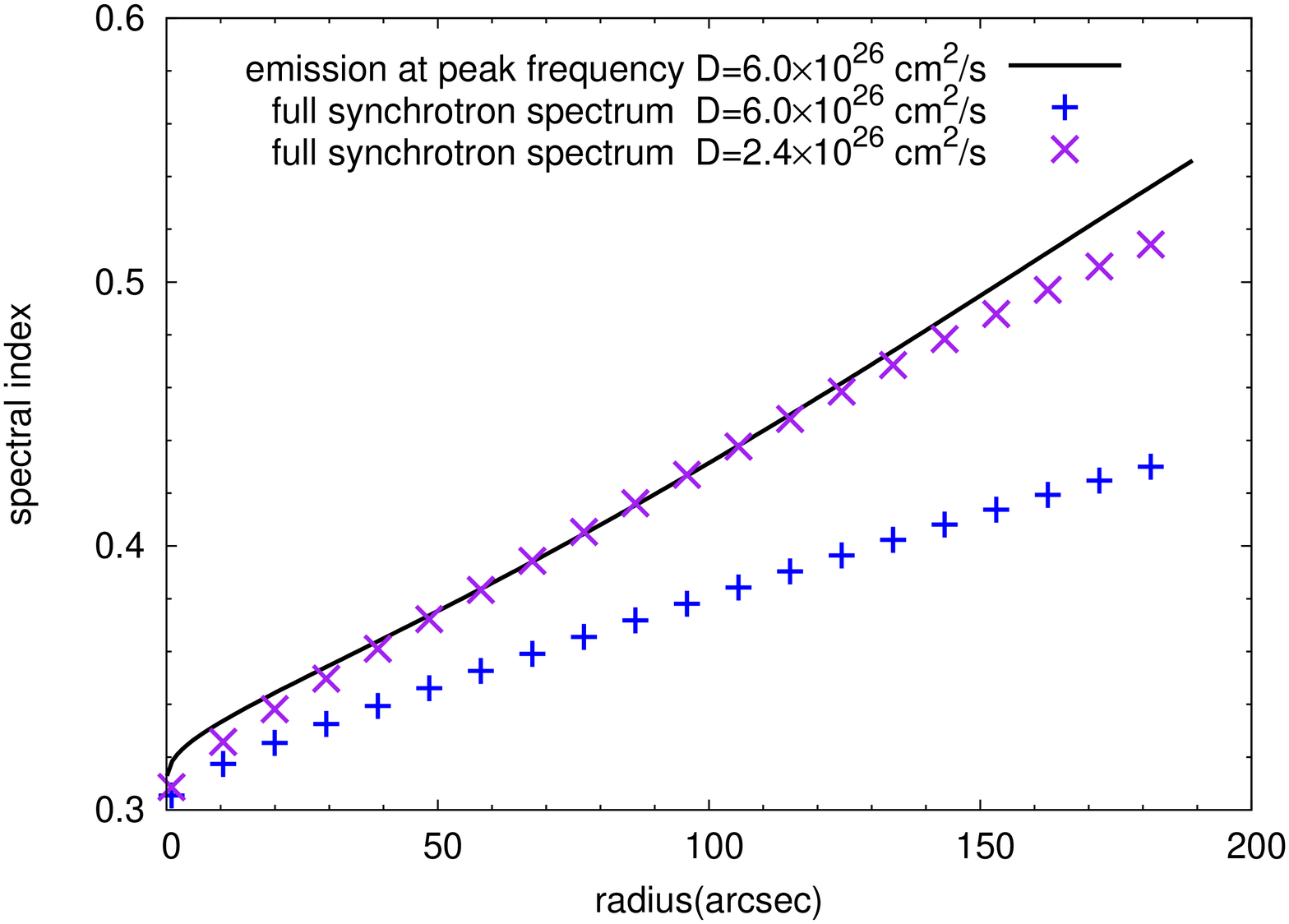} &
 \includegraphics[width=3.1in]{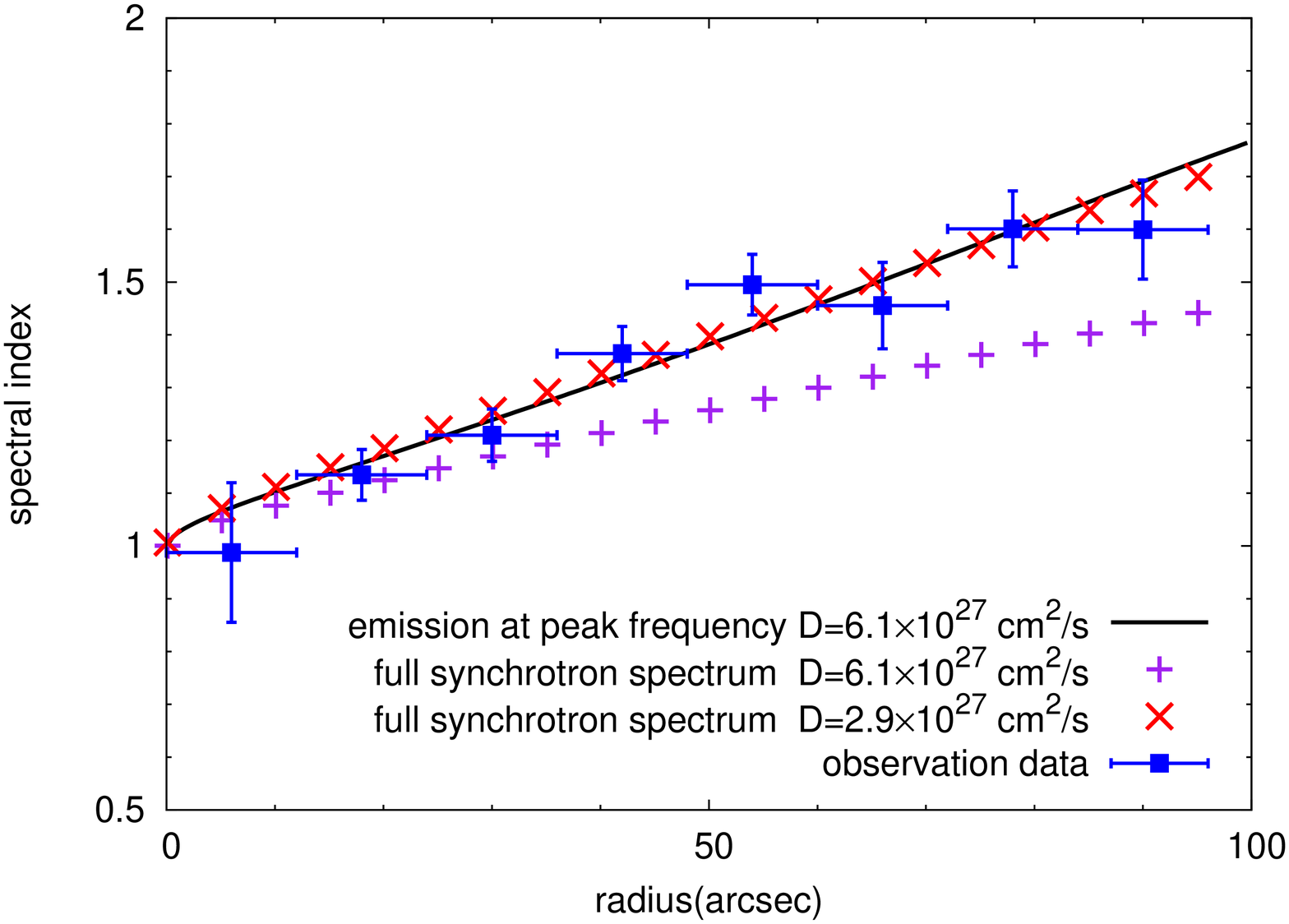} \\ 
(c)&\\
 \includegraphics[width=3.1in]{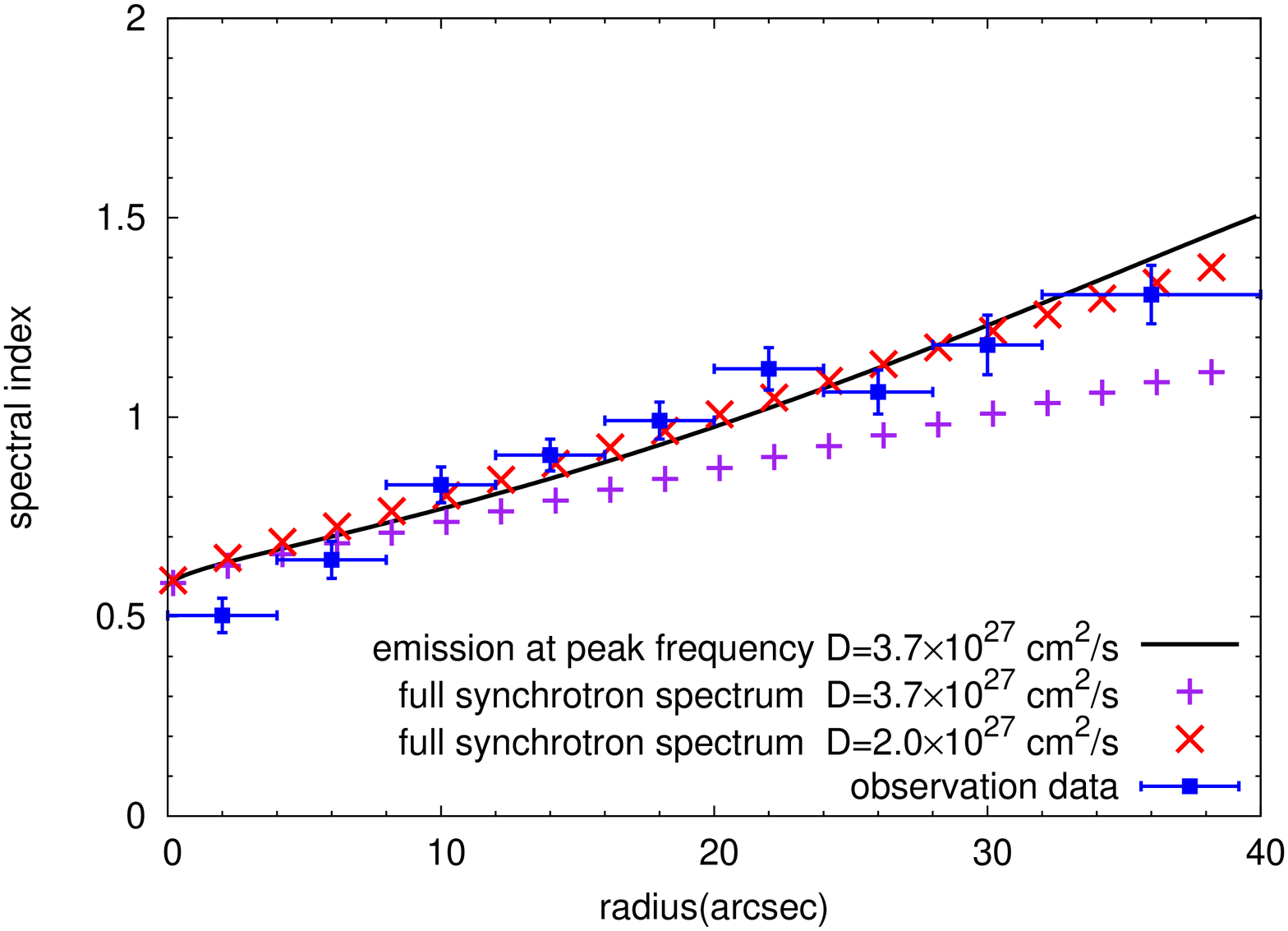} &
\end{array}$
\end{center}
\caption{\label{sync} (a) Flux spectral index ($\beta$) distribution of Crab from 5 GHz to $6\times 10^{14}$ Hz assuming $p_1=1.52$, $p_2=2.35$ and $B=300$ $\mu$G, (b) Flux spectral index ($\beta$) distribution of 3C 58 from 2.2 keV to 8 keV assuming $p=2.93$
and $B=80$ $\mu$G, (c) Flux spectral index ($\beta$) distribution of G21.5--0.9 from 0.5 keV to 10 keV assuming $p=2.08$
and $B=180$ $\mu$G.}
\end{figure*}

So far,  the results  are under the assumption that all the emission of an
electron goes into radiation at a frequency $\nu$ corresponding to the maximum
synchrotron radiation power. 
We will continue to make this assumption in the next section because it speeds the
calculations, but here we carry out the calculation with a full synchrotron spectrum
to show the uncertainty caused by  our approximation.
We considered the  Crab, 3C 58 and G21.5--0.9, 
and our numerical results with the full synchrotron spectrum  (Figure \ref{sync}) 
 show that the  diffusion coefficient $D$ required to fit the observations
drops by a factor about 2.  Here, we integrate the synchrotron radiation
function $F(x=\nu/\nu_c)$ \citep{pachol70} from 0.005 to 500 and assume that for 3C 58 and G21.5--0.9 all the particles within that energy range are in a steady state. 
Because $F(x=\nu/\nu_c)$ 
drops very quickly beyond the peak and we are calculating the spectral index distribution of 3C 58 and G21.5--0.9 in X-rays, time dependent effects are expected to be small.

\begin{figure}
        \epsscale{1.0} \plotone{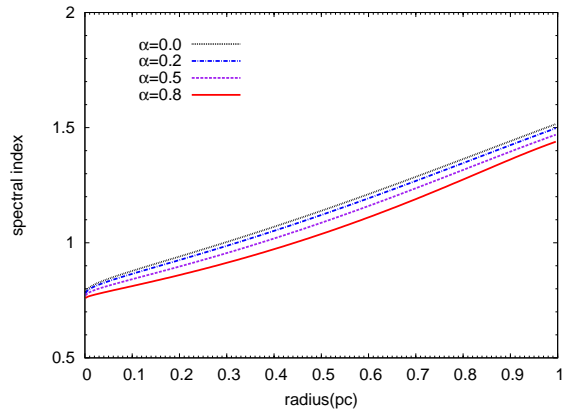}
        \caption{\label{DEfig} Flux spectral index ($\beta$) distribution at $10^{17}$ Hz for a PWN with $p_E=2.5-\alpha$, nebular size $R=1$ pc, magnetic field $B=100$ $\mu$G, and diffusion coefficient $D_E(\nu=10^{17}{\rm~Hz})=
10^{27}(1-\alpha)^2$ cm$^2$ s$^{-1}$, for various values of $\alpha$.
A steady state model with transmitting boundary is assumed.}

\end{figure}

In the pure diffusion model we have assumed that the diffusion coefficient is constant. 
However, it is possible that the diffusion coefficient in the PWN has energy dependence $D_E(E)$, as discussed in Section 2,
and we now investigate how the energy dependence of the diffusion coefficient would affect our pure diffusion model with transmitting boundary. We use the Green's function method to solve the steady state equation of the pure diffusion model but now with an energy dependent diffusion coefficient $D_E(E)=D_0E^\alpha$:
\begin{equation}
D_0E^{\alpha} \nabla^2N+Q\frac{\partial E^2 N}{\partial E}=-KE^{-p}\delta(\vec{r}).
\label{DEeq}
\end{equation}
The resulting particle distribution function is (see Appendix)  
\begin{eqnarray}
 N(E,r)=\frac{K}{4\pi r D_0E^\alpha}E^{-p}\frac{1}{\sqrt{\pi}}\nonumber\\ 
\times\int^\infty_u
\left(1-\frac{u}{x}\right)^{\frac{p+\alpha-2}{1-\alpha}}\frac{e^{-x}}{\sqrt{x}}dx,
\label{DEnum}
\end{eqnarray}
where
$$
u=\frac{r^2QE(1-\alpha)}{4D_0E^\alpha}.
$$
In the solution, $u$  changes to $u=r^2QE(1-\alpha)/4D_0E^\alpha$; again, $\sqrt u$ can be considered as a ratio of the diffusion distance to the nebular radius. In a steady state, $t=1/QE$,  and we have $\sqrt u=r(1-\alpha)^{1/2}/(4D_0E^\alpha t)^{1/2}$. Setting $u=1$, we find that  the nebular size $R\propto E^{-(1-\alpha)/2}\propto \nu^{-(1-\alpha)/4}$. For the spectral index distribution, we note that $\alpha>0$ implies that more energetic particles diffuse out more rapidly than less energetic particles, which should flatten the spectral index distribution; the same effect results from reducing the magnitude of the diffusion coefficient when $\alpha=0$.
Our simulations show that the case $D_E=D_0E^\alpha$ corresponds to a constant diffusion coefficient case with $D$ if $D_E=(1-\alpha)^2D$, where $D_E$ is taken at the energy at which the spectral index is measured.   
The flux spectral index at the center now is $(p_E+\alpha-1)/2$   (equation [\ref{DEnum}])  
because the integral part in the solution  approaches some constant when $r\rightarrow0$. In order to get the same spectral index at the center as in constant diffusion coefficient case we need $p_E=p-\alpha$.
In Figure \ref{DEfig},  we consider a PWN with  $p_E=p-\alpha=2.5-\alpha$, nebular size $R=1$ pc, magnetic field $B=100$ $\mu$G, and diffusion coefficient $D_E=10^{27} (1-\alpha)^2$ cm$^2$ s$^{-1}$ at an energy corresponding to $\nu=10^{17}$ Hz, and plot the spectral index profile for different $\alpha$ values. It is clear that they all have a similar slope, implying that for a certain band, if a pure diffusion model with constant $D$ can fit the data, a pure diffusion model with energy dependent $D_E$ can also fit the data but with different $p$ and diffusion coefficient at that band.  Although energy dependence of the diffusion coefficient does not change the consequences of  spectral index fitting with constant $D$ at one wavelength band, it changes the fitting result when multiband data are considered because now the nebular size $R\propto \nu^{-(1-\alpha)/4}$ instead of $R\propto \nu^{-1/4}$. We previously found with the constant diffusion coefficient assumption, there was a problem in explaining the nebular size of the Crab beyond optical frequencies. If we take the energy dependence of $D_E$ into account,  we can fit the data on nebular size and overall appearance of the spectral index profile in the IR and optical better. However 
it is difficult to determine whether energy dependence is required by the data. Since other  factors like advection  affect the nebular size and spectral index profile of the Crab in a similar way, and the observational data for the Crab Nebula  do not have high precision,  the energy dependence of the diffusion coefficient is  not determined by our models. 
However, the decreasing size at higher photon energies implies that $\alpha<1$; as discussed
in Section 2, some treatments of diffusion in old PWNe have assumed $\alpha=1$.
\subsection{Diffusion and advection model}

We now use Monte Carlo simulations to consider models with diffusion and advection. Monte Carlo methods allow the treatment of cases that cannot be treated  in the pure diffusion, analytical model. 
We expect that advection plays some role in the spectral index distribution and half light radius, especially in the $\nu \gg \nu_R$ regime. In the MHD model of KC, there is an advective flow after the termination shock in which the flow velocity declines from mildly relativistic to the velocity at the edge of the nebula,  $\sim 2000 \kms$.
If the magnetic field is not dynamically important, the velocity declines as $r^{-2}$ in a steady flow. 
Another consideration is a  reflecting outer boundary condition. For PWNe satisfying $\eta=(6Dt/R^2)^{1/2}\gg 1$, the reflecting boundary should have a significant effect on the spectral index distribution
and half light radius. For the reflecting boundary effect we primarily focus on 3C 58, which has a high diffusion coefficient $D$. The last improvement is treating an extended source, since the termination shock has a finite size  $\sim 0.1$ pc which can play a role in the central region. The time dependence  of the PWN and non-spherical symmetry are  not taken into account.

\cite{massaro85} discussed  a diffusion and advection model, but many assumptions were needed to obtain an analytical model. 
\cite{weinberg76} and \cite{reynolds91} investigated  approximate solutions for a pure diffusion model with an extended source, but they omit some of the physical effects that we wish to include.     
We therefore developed a code to carry out Monte Carlo simulations that allowed a general
treatment of diffusion and advection processes. We modeled both the Crab and 3C 58. For 3C 58 we mainly focus on the variation of spectral index distribution while for the Crab we are interested in the variation of the half light radius with photon energy. In simulating 3C 58, a total number of $10^6$ effective particles were injected into a spherical shell  at time intervals of $5.4\times 10^4$ s. Since we are considering the X-ray band of 3C 58, which is in a steady state,  only particles injected within a time $1/QE$ of the present  are taken into account. In simulating the Crab, at least $10^6$ effective particles were injected into a spherical shell at time intervals of a half day. The motion of the particles is a superposition of advective motion and 3-dimensional random motion. 
The displacement in each time interval is
\begin{eqnarray}
\Delta \vec{S}_{tot}=V_{adv}\Delta t \,\vec{r}_o\pm(2D_x\Delta t)^{1/2}\,\vec{x}_o \nonumber \\
\pm(2D_y\Delta t)^{1/2}\,\vec{y}_o\pm(2D_z\Delta t)^{1/2}\,\vec{z}_o, 
\end{eqnarray}
where $\vec{x}_o,\vec{y}_o,\vec{z}_o$ and $\vec{r}_o$ are unit vectors in $x,y,z$ and radial directions.
We take $V_{adv}\propto r^{-2}$, $V(r=R_{out})=630\kms$\citep{bieten06} and the ratio of outer radius to inner radius to be $R_{out}/R_{in}=100/8=12.5$ for 3C 58 \citep{slane04}. In this case, we only consider the synchrotron radiation loss since the adiabatic expansion energy loss, $\dot E\propto \nabla \cdot { V_{adv}}$,  is zero for our
velocity profile. For the Crab, we use some of the parameters from the model of \cite{kennel84b}, assuming that the velocity at the termination shock is $c/3$, the velocity $V_{adv}\propto r^{-2}$ between the termination shock radius $R_s$ and a critical radius $R_c$, and $V_{adv}$ remains constant between the critical radius $R_c$ and the outer radius of the nebula $R_{out}$. We assume $R_s=15$ arcsec in angular size \citep{hester02}, and in order to set $V_{adv}(R_{out})=2000$ km s$^{-1}$ \citep{kennel84a} we further assume $R_c=50^{1/2}R_s$. We consider  adiabatic expansion energy losses in addition to synchrotron radiation losses for $R_c<R<R_{out}$; it is $dE/dt=-2V_{adv}(R_{out})E/3r$ for relativistic particles in a medium with constant flow velocity. In all the simulations, we use a constant magnetic field $B$ to simplify the calculation.  
In spherical symmetry, $D_x=D_y=D_z\equiv D$.   
If particles move into the inner boundary
due to their random motions, they are forced to bounce back into the PWN region. 
A reflecting boundary was used at the outer radius. 
 
\begin{figure*}
\begin{center}
$\begin{array}{cc}
(a)&(b)\\
  \includegraphics[width=3.1in]{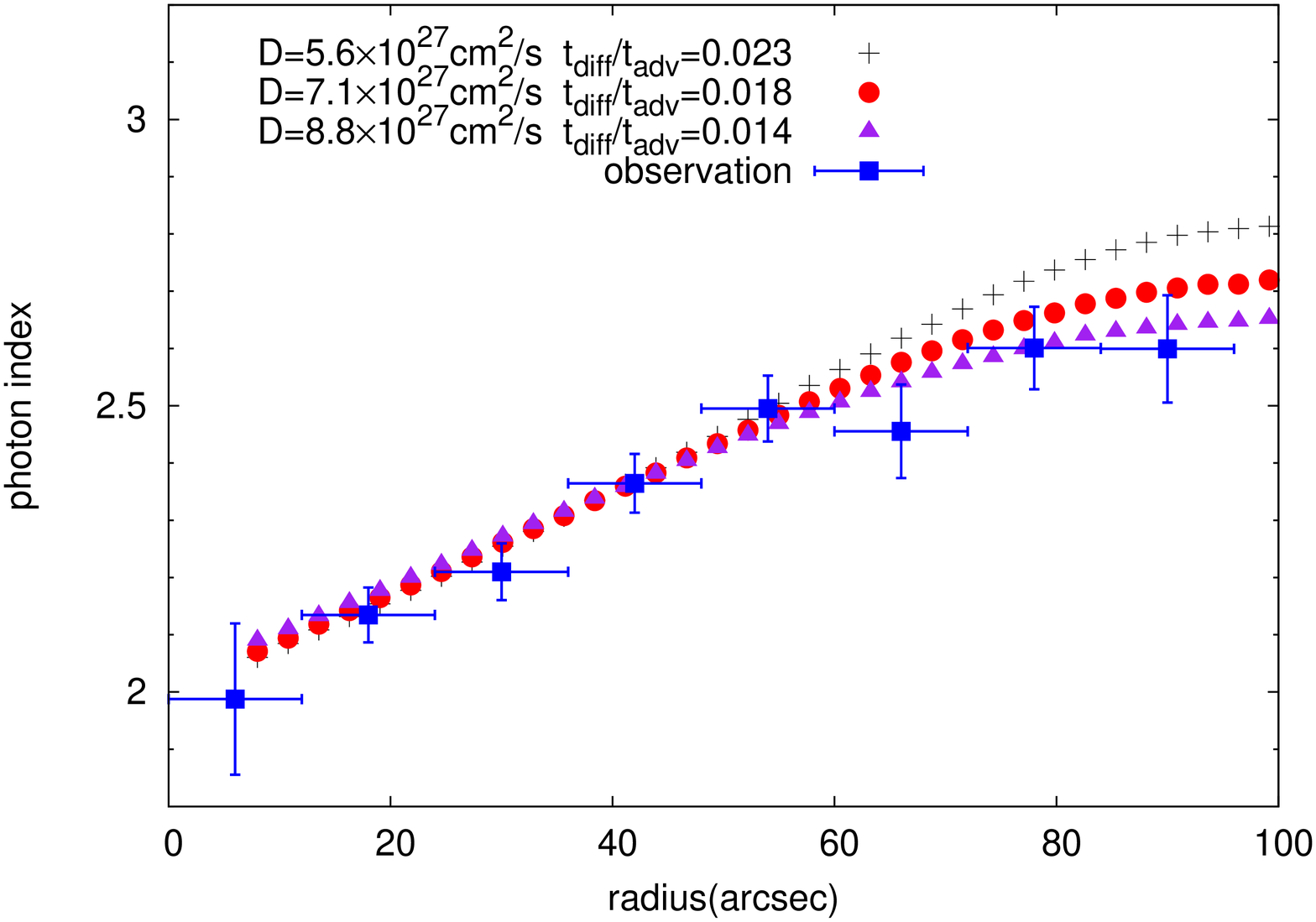} &
  \includegraphics[width=3.1in]{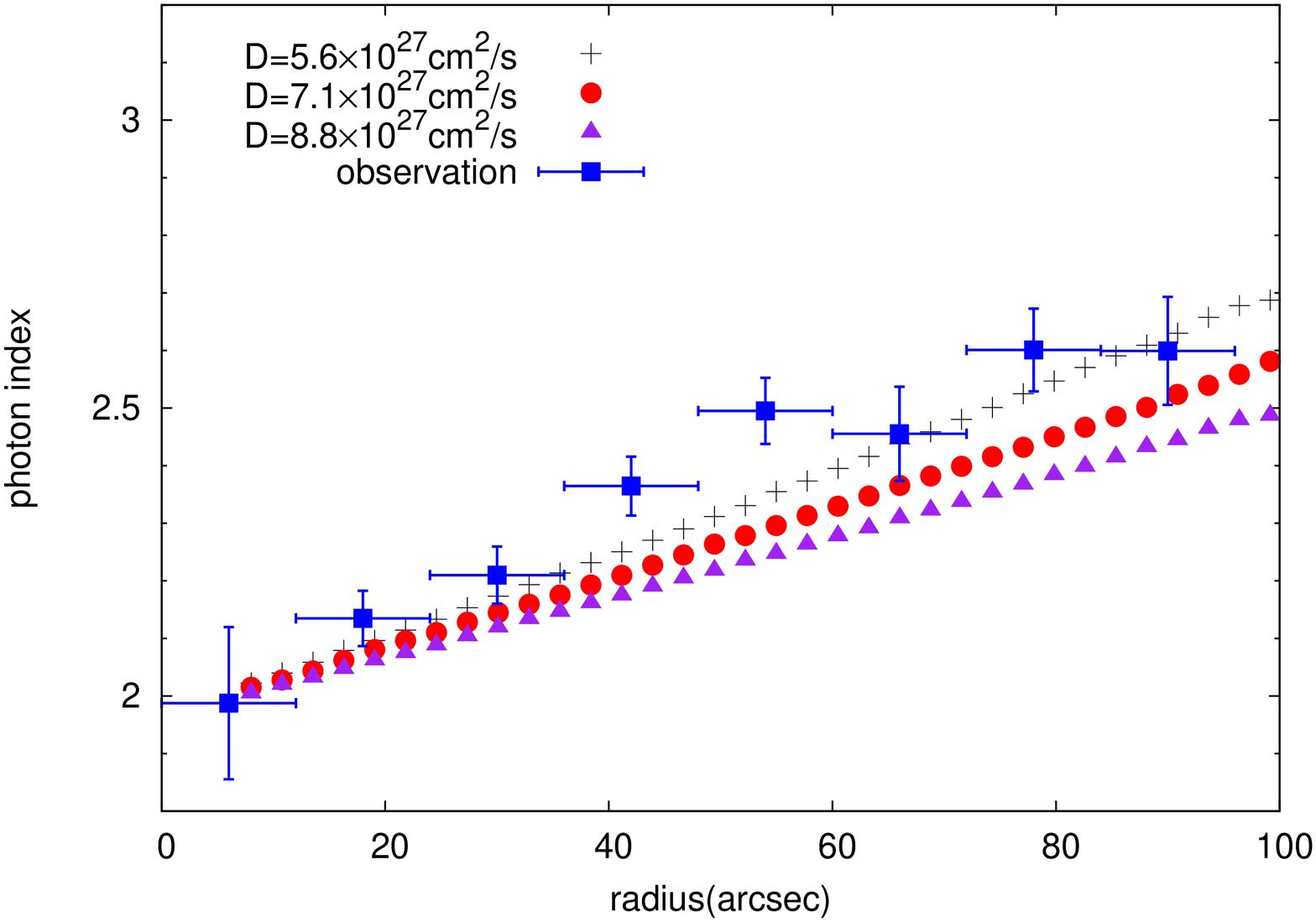} \\ 
(c)&\\
  \includegraphics[width=3.1in]{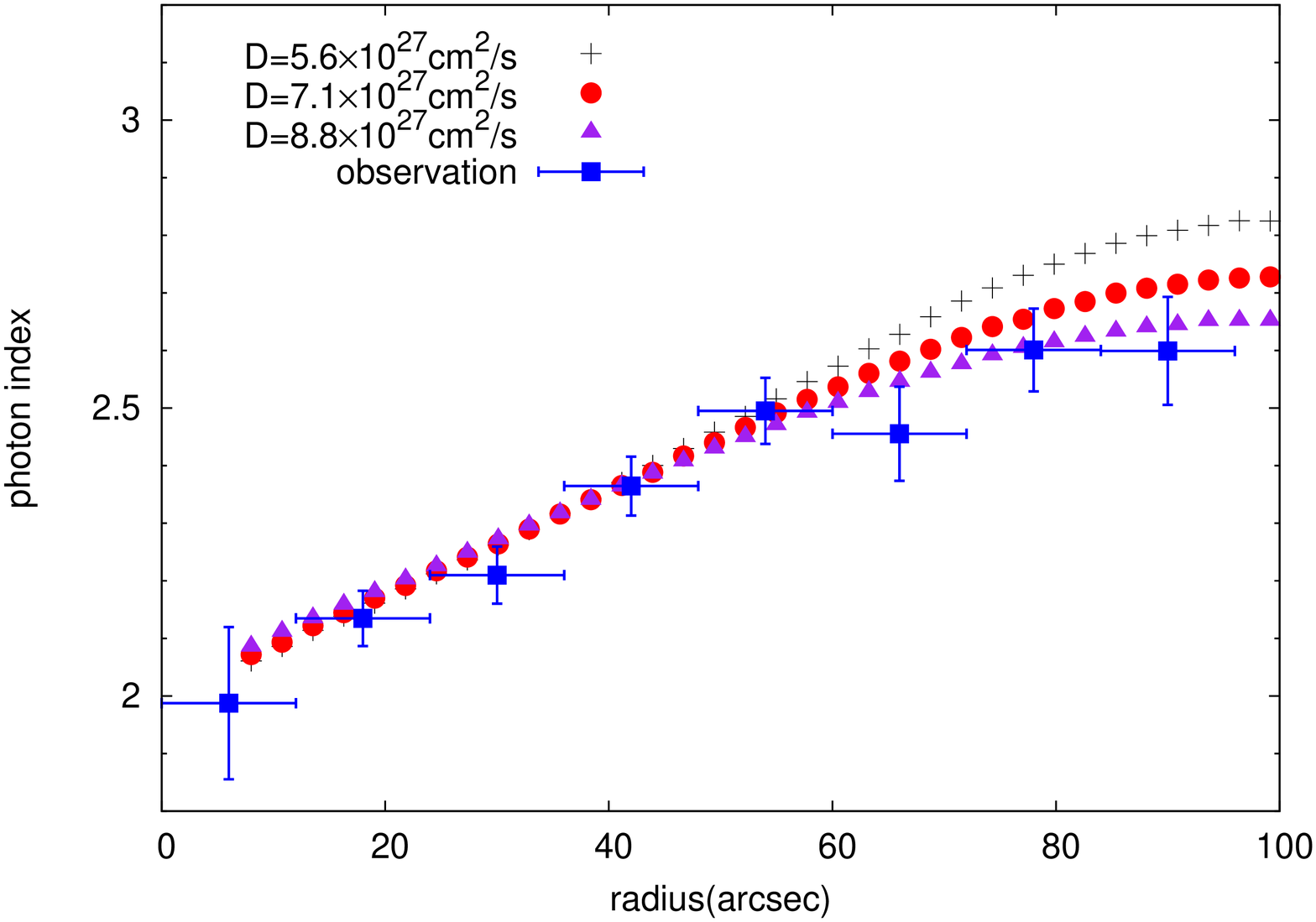} &\\
\end{array}$
\end{center}
\caption{\label{Montecarlo} (a) Monte Carlo simulation with diffusion and advection for 3C 58 from 2.2 keV to 8 keV assuming $p=2.8$, $B=80\mu G$, and reflecting inner and outer boundaries, (b) Monte Carlo simulation with pure diffusion, a reflecting inner boundary, and a tranmitting outer boundary, (c) 
Monte Carlo simulation with pure diffusion and reflecting inner and outer boundaries.}
\end{figure*}

The simulation results for 3C 58 are shown in Figure \ref{Montecarlo}. The additional physical processes allow a better fit to the data. In order to discern what role advection and a reflecting boundary  play in the photon index distribution, we did  one simulation for pure diffusion with only an inner reflecting boundary and one for pure diffusion with both inner and outer reflecting boundaries (Figure \ref{Montecarlo}). After comparing the results of the two cases (Figure \ref{Montecarlo}),  we find that the outer reflecting boundary condition mainly makes the photon index  steeper and the part near the outer boundary relatively flat. The  diffusion coefficient $D$ required to fit the data  becomes higher. Advection is not very important in the fit because the ratio of diffusion time to advection time ratio is low. In the Figures
\ref{Montecarlo}, we have shown the ratio of diffusion time to advection time  $t_{diff}/t_{adv}$, where $t_{diff}$ is estimated by $(6Dt_{diff})^{1/2}=R_{out}-R_{in}$ and $t_{adv}$ is calculated by $t_{adv}=\int^{R_{out}}_{R_{in}}dr/V_{adv}$. 
The angular size of the flat region in the radial direction can be estimated as follows. Since $t_{diff}\ll t_{adv}$ in our simulation,
 advection can be neglected. For particles in a steady state, the diffusion distance is 
$R_{diff}=(6Dt)^{1/2}=(6D/QE)^{1/2}$, so that $\theta_{flat}\approx\theta\times (R_{diff}/R-1)$, where $\theta$ is the angular size of PWN.   Substituting the expression for $R_{diff}$  and recalling that $E_R$ satisfies $R=(4D/QE_R)^{1/2}$, we obtain  
\begin{eqnarray}
 \theta_{flat}\approx\theta \left(\left[\frac{3E_R}{2E}\right]^{1/2}-1\right)\nonumber\\
=\theta \left(\frac{6^{1/2}}{2}\left[\frac{\nu_R}{\nu}\right]^{1/4}-1\right).
\label{Montecarlo_flat}
\end{eqnarray}
For 3C 58 with $D=8.8\times 10^{27}$ cm$^2$ s$^{-1}$, $B=80$ $\mu$G, and $R=1.55$ pc,  equation (\ref{Montecarlo_flat}) implies $\theta_{flat}\approx 30^{\prime\prime}$ which is consistent with the results in Figure \ref{Montecarlo}. The results shown in Figure \ref{Montecarlo} are calculated for $2.2- 8$ keV
photon energies. When we do the calculation for $\theta_{flat}$ we use 8 keV as it gives a lower diffusion distance. We  emphasize that equation (\ref{Montecarlo_flat}) is only correct for a steady state and requires that $1\gg \frac{\sqrt{6}}{2}(\frac{\nu_R}{\nu})^{\frac{1}{4}}-1>0$. When $\frac{\sqrt{6}}{2}(\frac{\nu_R}{\nu})^{\frac{1}{4}}-1<0$, the cooling time is lower than the diffusion time, and few particles  reach the outer boundary. When $\frac{\sqrt{6}}{2}(\frac{\nu_R}{\nu})^{\frac{1}{4}}-1>1$, due to the boundary effect, the diffusion coefficient $D$ is large enough to smear out the spectral index structure in the nebula, so the spectral index distribution is flat within that energy band. 

\begin{figure}
        \epsscale{1.0} \plotone{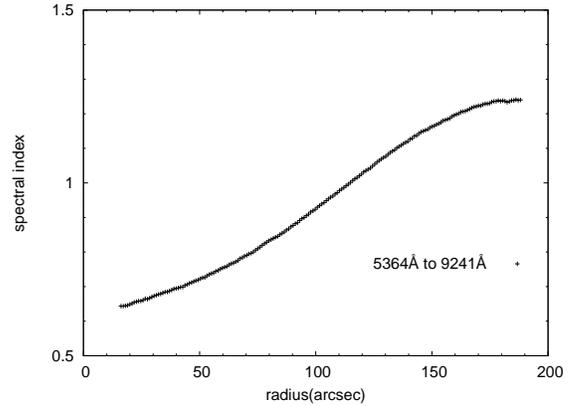}
        \caption{\label{crab_Montecarlo} {Flux spectral index ($\beta$) distribution of the Crab at optical wavelengths assuming $p=2.0$, $B=300$ $\mu$G, and $D=9\times 10^{26}$ cm$^2$ s$^{-1}$ in a Monte Carlo simulation with reflecting inner and outer boundaries.}}
\end{figure}

We calculate the spectral index distribution of the Crab at optical wavelengths (Figure \ref{crab_Montecarlo}) and its half light radius  from ultraviolet (UV) to X-ray frequencies (Table \ref{crab_montecarlo_radius}), which are above the break frequency given by \cite{buccian11}, with $p_2=2.0$, $B=300$ $\mu G$ and $D=9.0\times 10^{26}$ cm$^2$ s$^{-1}$. The spectral index variation from 0.6 to 1.1 within the Crab Nebula at optical wavelengths \citep{veron93} now implies the nebular size of Crab is  $\sim140^{\prime\prime}$, which is slightly larger than the pure diffusion case as we now use a larger diffusion coefficient $D$ and take advection into account.    The advection process increases the half-light radius of the Crab significantly at X-ray energies. Based on Table \ref{crab_montecarlo_radius}, the ratio $(R_{diff}-R_s)/(R_{diff+adv}-R_s)$  increases rapidly from X-ray to UV wavelengths as we assume the extended source size  is $\sim15^{\prime\prime}$. The reason  advection is important for the Crab in X-rays is because the X-ray size of the Crab is small. As noted in Section 3.1, the ratio of advection time to diffusion time $R_{adv}/R_{diff} \propto R$, the nebular size; advection becomes more important when the nebula  is small. The half-light radius derived from the diffusion and advection model can now explain the high frequency part of the Crab nebula size data  \citep{amato00}.  However, the half-light radius we obtained still drops  sharply as a function of frequency, which may imply  energy dependent diffusion. The values of $p_2$, magnetic field $B$, and diffusion coefficient $D$ we choose are not best fit parameters. Certain combinations of parameters would improve the fit to  the observational data. The velocity and magnetic field profiles for the diffusion and advection case must be analyzed for more exact models. The uncertainty in the termination shock radius and the flow velocity at the outer radius of the nebula  also affect the simulation results. 
\begin{table}
\centering
\caption{Half light radius of the Crab Nebula in arcsec}
\begin{tabular}{lcccc}
\hline\hline
 Frequency &$10^{15}$ Hz&$10^{16}$ Hz&$10^{17}$ Hz&$10^{18}$ Hz \\ \hline
pure diffusion & 72 & 43 & 27 & 19\\ \hline
diffusion and advection& 84& 55& 35 & 24\\ \hline
\end{tabular} 
\label{crab_montecarlo_radius}
\end{table}

As discussed in Section 1, 2-dimensional MHD models reproduce many features observed in
the inner Crab Nebula.
Diffusion is not a factor in this region for 2 reasons: the short advection timescale
because of the high flow velocities and a long timescale for radial diffusion because
of the toroidal magnetic field.
In the Crab, the prominent toroidal structure observed at X-ray and optical wavelengths
extends to $\sim 40\arcsec$ from the pulsar, while the nebular radius is $200\arcsec$.
Our model applies to the outer 4/5 of the nebula.
Toroidal structure is less prominent in 3C 58 and G21.5--0.9.
   
We also used Monte Carlo simulations to investigate the case where there is particle transport across
the nebular boundary at $R$ and particles are lost from the system once they cross $R$.
The effect is to flatten the spectral index profile in the outer part of the nebula and to reduce
the value of $D$ by a factor of 2 compared to the simple model (Section 3.1).
However, we consider the reflecting boundary model to be more plausible, for the
reasons given in Section 2.

\section{DISCUSSION AND CONCLUSIONS}

We have argued that the structure of young PWNe is not described by a toroidal
magnetic field, as expected in a model like that of \cite{kennel84a}, but has a more
chaotic magnetic structure.
In Section 2, we noted various observational studies that showed considerable
structure in young nebulae, not a clear toroidal structure.
In Section 3, models with diffusion of particles were presented and compared to
observations of 3 nebulae.
Emphasis was placed on fitting the spectral index profiles of the nebulae, as well as
the surface brightness profiles.
Models with diffusion were much better able to spectral index profiles than pure
advection models.
The best estimates of the diffusion coefficient come from the Monte Carlo simulations,
but these values need to be somewhat reduced because they do not include the full
synchrotron spectrum in the calculation of the emission.
Estimates of the diffusion coefficient and the corresponding particle mean free path
are given in Table \ref{diffusion}.
The assumed magnetic field is given for each case because $D\propto B^{3/2}$ and the
magnetic field strength is uncertain.

Table \ref{diffusion} shows that the diffusion coefficient and length for the Crab is
considerably smaller than that for 3C 58 and G21.5--0.9. 
The length does not scale with the size of the PWN because the Crab is 
larger than both 3C 58 and G21.5--0.9.
One possibility is that there is frequency dependent diffusion coefficient.
The coefficient for the Crab is derived from optical/IR observations, so that
the lower energy particles are being observed compared to 3C 58 and G21.5--0.9
where X-ray observations are used.
Table \ref{diffusion}  shows the corresponding particle energies for diffusion coefficients.
An energy dependent diffusion with $\alpha\sim 0.5-0.6$ would explain the difference
between the Crab and the other remnants.
As discussed in Section 3.1, this is consistent with our results and we suggest it as
a possible explanation for the difference between the PWNe.
\begin{table}[ht]
\centering
\caption{Diffusion coefficient and length}
\begin{tabular}{lcccc}
\hline\hline
 Object &  $D$  &  $B$ &  $\lambda=D/c$ & Particle energy \\ 
  &   (cm$^2$/s) &   ($\mu$G) &   ($10^{16}$ cm)&    (TeV) \\ \hline
Crab & $2.4\times 10^{26}$  & 300 & 0.8 &    0.6 \\ \hline
3C 58 & $2.9\times 10^{27}$   & 80 & 10   &    40   \\ \hline
G21.5-0.9&  $2.0\times 10^{27}$  & 180 &  7  &   30   \\ \hline

\end{tabular} 
\label{diffusion}
\end{table}

The spectral indices along magnetic filaments in the Crab shows relatively little
variation while the spectral indices show steepening away from the filament
center \citep{seward06,hester08}, which is consistent with rapid particle motions
along filaments and slow diffusion across filaments.
However, the diffusion mean free paths (Table \ref{diffusion}) are smaller than characteristic
structures in the nebulae.
The length for the Crab is about a factor 10 smaller than the scale indicated by
optical polarization (Section 2), and the length for 3C 58 is about a factor of 10 smaller
than the scale of apparent magnetic loops seen in the X-ray image \citep{slane04}.
The actual longer diffusion time (due to the smaller length) may indicate that the
magnetic structure is not completely random. 

Our models have been designed for comparison with young PWNe like the Crab, 3C 58,
and G21.5--0.9.
They are likely to be in an evolutionary phase where the nebulae are accelerating into
freely expanding supernova ejecta.
In a subsequent phase of evolution, the reverse shock wave from interaction with the
interstellar medium comes back toward the center and can push off the PWN,
creating an asymmetric nebula \citep{blondin01}.
\cite{etten11} have investigated evolutionary models for the PWN HESS J1825--137,
which probably belongs to the class of post-reverse shock nebulae and is
 observed at X-ray and TeV energies.
Of interest is the fact that their modeling shows the need for diffusion of particles.
As mentioned in Section 2, \citet{hinton11} have considered diffusion from the evolved
PWN Vela X.
A chaotic magnetic field is expected in these objects because of instabilities related to
the interaction with the reverse shock front from the supernova remnants \citep{blondin01}.
The situation is different for the young remnants discussed here, but a chaotic field
may be the result of 
instabilities related to the toroidal magnetic field \citep{begel98} and Rayleigh-Taylor instabilities in the outer parts of the nebulae.

\acknowledgments
We are grateful to the referee for helpful comments.
This research was supported in part by
NASA grant NNX07AG78G and NSF grant AST-0807727.

\section*{Appendix}
In order to solve the equation
\begin{equation}
D_0E^{\alpha} \nabla^2N+Q\frac{\partial E^2 N}{\partial E}=-KE^{-p}\delta(\vec{r}),
\label{DE}
\end{equation}
we first let $\Phi=QE^2N$ and $4\pi J=\frac{Q}{D_0}E^{2-\alpha}KE^{-p}\delta(\vec{r})$, so equation  (\ref{DE}) can be simplified to 
\begin{equation}
 \nabla^2\Phi -\frac{Q}{D_0}(1-\alpha)\frac{\partial \Phi}{\partial E^{\alpha-1}}=-4\pi J.
\end{equation}
Further, assuming $m^2=\frac{Q}{D_0}(1-\alpha)$ and $\tau=E^{\alpha-1}$, we have
\begin{equation}
 \nabla^2\Phi -m^2\frac{\partial \Phi}{\partial \tau}=-4\pi J.
\end{equation}	
Here we only consider the case $m^2>0$, which requires $\alpha <1$, as it is consistent with the 
situation in the Crab. If $\alpha > 1$, the nebular size of the Crab does not 
decrease with increasing frequency. 
Next we consider the Green's function equation corresponding to the above equation:
\begin{equation}
 \nabla^2G -m^2\frac{\partial G}{\partial \tau}=-4\pi \delta(\tau-\tau') \delta(\vec{r}-\vec{r'}).
\label{greeneq}
\end{equation}
If we do not consider the boundary effect, the Green's function solution for equation (\ref{greeneq}) is 
\begin{equation}
 G(r,\tau/r',\tau')=\frac{m}{2\sqrt{\pi}}\frac{1}{(\tau-\tau')^{1.5}}e^{-m^2\frac{|r-r'|^2}{4(\tau-\tau')}}u(\tau-\tau'),
\end{equation}
 where $u(\tau-\tau')$ is the step function.  Here, $G(r,\tau/r',\tau')$ is already normalized to 
\begin{equation}
\frac{m^2}{4\pi}\int Gd\vec{V}=1.
\end{equation}
Then
\begin{equation}
 \Phi(r,\tau)=\int^\tau_0 d\tau' \int d\vec{V'} J(r',\tau')G(r,\tau/r',\tau').
\end{equation}
After some calculation, we obtain 
\begin{equation}
 \Phi(r,E)=K\frac{Q}{D_0}\frac{1}{4\pi^{1.5}r}E^{-p-\alpha+2}\int^\infty_u \left(1-\frac{u}{x}\right)^\frac{p+\alpha-2}{1-\alpha}\frac{e^{-x}}{\sqrt{x}}dx,
\end{equation}
where $u=\frac{Q(1-\alpha)r^2}{4D_0}E^{1-\alpha}$. Since $N(r,E)= \Phi/QE^2$, we finally obtain 
\begin{equation}
 N(r,E)=\frac{K}{4\pi r}\frac{E^{-p}}{D_0 E^\alpha}\int^\infty_u\frac{1}{\sqrt{\pi}} \left(1-\frac{u}{x}\right)^\frac{p+\alpha-2}{1-\alpha}\frac{e^{-x}}{\sqrt{x}}dx.
\end{equation}


\end{document}